\documentclass[journal,hidelinks]{IEEEtran}

\usepackage{graphicx,url}
\usepackage[english]{babel}
\addto\captionsenglish{}
\usepackage[utf8]{inputenc}
\usepackage{amsmath}
\usepackage{amssymb}
\usepackage{mathtools}
\usepackage{cuted}
\usepackage{cite}
\usepackage{balance}
\usepackage{tikz}
\usepackage{pgfplots}
\usepackage{tikzscale}
\pgfplotsset{compat=newest}
\usepgfplotslibrary{groupplots}
\usetikzlibrary{arrows.meta, positioning, calc}
\newlength{\figurewidth}
\newlength{\figureheight}
\definecolor{myblue}{rgb}{0.0000,0.4470,0.7410}
\definecolor{myyellow}{rgb}{0.9290,0.6940,0.1250}
\definecolor{myred}{rgb}{0.6350,0.0780,0.1840}
\definecolor{mygreen}{rgb}{0.4660,0.6740,0.1880}
\definecolor{mypurple}{rgb}{0.4940,0.1840,0.5560}
\definecolor{myorange}{rgb}{0.8500,0.3250,0.0980}
\definecolor{mycyan}{rgb}{0.3010,0.7450,0.9330}
\usepackage{scalefnt}
\usepackage{orcidlink}
\usepackage[acronym,shortcuts]{glossaries}
\usepackage{algorithm, algpseudocode}
\usepackage{algcompatible}
\usepackage{bm}
\usepackage{booktabs}

\newacronym{RMSE}{RMSE}{root mean square error}
\newacronym{MMSE}{MMSE}{minimum mean square error}
\newacronym{MF}{MF}{matched filter}
\newacronym{OTFS}{OTFS}{orthogonal time frequency space}
\newacronym{AFDM}{AFDM}{affine frequency division multiplexing}
\newacronym{ISAC}{ISAC}{integrated sensing and communications}
\newacronym{QAM}{QAM}{quadrature amplitude modulation}
\newacronym{AWGN}{AWGN}{additive white Gaussian noise}
\newacronym{OFDM}{OFDM}{orthogonal frequency division multiplexing}
\newacronym{DFT}{DFT}{discrete Fourier transform}
\newacronym{IDFT}{IDFT}{inverse discrete Fourier transform}
\newacronym{TD}{TD}{time-domain}
\newacronym{CP}{CP}{cyclic prefix}
\newacronym{CPP}{CPP}{chirp-periodic prefix}
\newacronym{DAFT}{DAFT}{discrete affine Fourier transform}
\newacronym{IDAFT}{IDAFT}{inverse discrete affine Fourier transform}
\newacronym{ICI}{ICI}{inter-carrier interference}
\newacronym{ISI}{ISI}{inter-symbol interference}
\newacronym{BER}{BER}{bit error rate}
\newacronym{iid}{i.i.d.}{independent and identically distributed}
\newacronym{IoT}{IoT}{internet of things}
\newacronym{V2X}{V2X}{vehicle-to-everything}
\newacronym{LEO}{LEO}{low-earth orbit}
\newacronym{DD}{DD}{doubly-dispersive}
\newacronym{6G}{6G}{sixth generation}
\newacronym{CSI}{CSI}{channel state information}
\newacronym{PLS}{PLS}{physical-layer security}
\newacronym{QPSK}{QPSK}{quadrature phase-shift keying}
\newacronym{LMMSE}{LMMSE}{linear minimum mean square error}
\newacronym{SNR}{SNR}{signal-to-noise ratio}
\newacronym{OOBE}{OOBE}{out-of-band emission}
\newacronym{PAPR}{PAPR}{peak-to-average power ratio}
\newacronym{AFBM}{AFBM}{affine filter bank modulation}
\newacronym{FBMC}{FBMC}{filter bank multicarrier modulation}
\newacronym{PPN}{PPN}{polyphase network}
\newacronym{SIR}{SIR}{signal-to-interference ratio}
\newacronym{SINR}{SINR}{signal-to-interference-plus-noise ratio}
\newacronym{CCDF}{CCDF}{complementary cumulative distribution function}
\newacronym{GAS}{GAS}{Grover adaptive search}
\newacronym{5G}{5G}{fifth generation}
\newacronym{4G}{4G}{fourth generation}

\newcommand{\herm}[0]{^{\mathsf{H}}}
\newcommand{\Exp}[0]{\mathbb{E}}

\begin{document}

\title{Physical Layer Security in Chirp-Permuted AFBM}

\author{Pedro Schuves Marodin\textsuperscript{\orcidlink{0009-0007-6418-4696}}, \IEEEmembership{Student Member,~IEEE,}
Henrique L. Senger\textsuperscript{\orcidlink{0009-0004-1586-8168}},
Gustavo P. Gonçalves\textsuperscript{\orcidlink{0009-0000-8260-4390}},\\
Hyeon Seok Rou\textsuperscript{\orcidlink{0000-0003-3483-7629}}, \IEEEmembership{Member,~IEEE,}
Bruno S. Chang\textsuperscript{\orcidlink{0000-0003-0232-7640}}, \IEEEmembership{Member,~IEEE,}\\
Giuseppe Thadeu Freitas de Abreu\textsuperscript{\orcidlink{0000-0002-5018-8174}}, \IEEEmembership{Senior Member,~IEEE,}\\ and Didier Le Ruyet \IEEEmembership{Senior Member,~IEEE,}

\thanks{H~S.~Rou and G.~T.~F.~de~Abreu are with the School of Computer Science and Engineering, Constructor University, Campus Ring 1, 28759 Bremen, Germany (emails: [hrou, gabreu]@constructor.university).}
\thanks{P.~S.~Marodin, H.~L.~Senger, G.~P.~Gonçalves and B.~S.~Chang are with the CPGEI/Electronics Department, Federal University of Technology - Paraná, Curitiba, Brazil (emails: [pmarodin, hsenger, gustavog.1999]@alunos.utfpr.edu.br, bschang@utfpr.edu.br).}
\thanks{D.~Le~Ruyet is with the LAETITIA/CEDRIC/Conservatoire National des Arts et Métiers - Paris, France (email: didier.le$\_$ruyet@cnam.fr).}
\vspace{-5ex}
}

\maketitle

\begin{abstract}
We present a physical-layer secure communications scheme based on a double chirp-permuted variant of the affine filter bank modulation (AFBM) waveform, recently proposed for sixth-generation (6G) and integrated sensing and communications (ISAC) systems. Exploiting the dual-transform architecture of AFBM, utilized for frequency spreading and precoding, the scheme binds a secret permutation key to the chirp sequence that each transform interposes between two Fourier stages. A mismatch in either key is conjugated by a discrete Fourier transform and becomes a dense rather than a diagonal product, so that the complex orthogonality restoration on which AFBM depends fails and the filter bank's intrinsic interference leaks into the detector. Over a doubly-dispersive channel, an eavesdropper holding either key wrong cannot detect reliably and obtains no usable rate. A blind carrier-phase attack does not improve this, even with perfect channel state information (CSI) and co-location with the legitimate receiver. A single-permutation variant, in which one key permutes the same chirp in both transforms, is also examined.
\end{abstract}


\IEEEpeerreviewmaketitle

\glsresetall

\vspace{-3ex}
\section{Introduction}
The transition to sixth-generation (6G) networks demands reliable communication in highly dynamic environments such as vehicle-to-everything (V2X) and low-earth orbit (LEO) satellite links, with integrated sensing and communications (ISAC) as a defining feature. These scenarios involve doubly-dispersive (DD) channels with multipath and large Doppler shifts \cite{rou2024from}, in which orthogonal frequency division multiplexing (OFDM) suffers from inter-carrier interference (ICI). Orthogonal time frequency space (OTFS) and affine frequency division multiplexing (AFDM) reach full diversity in DD channels \cite{bemani2023affine}. Affine filter bank modulation (AFBM) adds a polyphase filter bank to the discrete affine Fourier transform (DAFT) spreading of AFDM to improve spectral containment. Since filter bank multicarrier (FBMC) systems lack complex orthogonality, AFBM uses a precoder that restores it and removes the intrinsic interference~\cite{ranasinghe2026afbm, senger2025afbm_waveform}.

The broadcast nature of these networks makes them open to eavesdropping, and upper-layer cryptography adds computation and latency that conflict with Internet of Things (IoT) and ISAC requirements. Physical-layer security (PLS) instead draws on the channel and on the waveform itself~\cite{mucchi2021pls_6g}. Against passive eavesdroppers that demodulate by brute force, AFDM reaches a bit error rate (BER) of 0.5 by keeping its chirp parameters secret, at a search cost of $O(N^2)$ for $N$ subcarriers~\cite{savaux2026robustness}. Chirp-permuted AFDM (CP-AFDM) permutes the chirp sequence, which raises the search space to $N!$ and resists quantum search~\cite{rou2025chirppermuted, rou2025cpafdm_versatile}, but it does not address spectral containment. Other secure designs rely on frequency agility~\cite{boroujeni2025rfpa} or on the AFDM parameters~\cite{zhang2026secure} and do not use a permutation of the waveform structure.

AFBM is a suitable host for such a permutation. Its prototype filter confines the spectrum far below the rectangular-window envelope of AFDM, reaching $-100$~dB of out-of-band emission (OOBE) when a long prototype is admissible~\cite{ranasinghe2026afbm, senger2025afbm_waveform}, and with a Gaussian belief propagation receiver it outperforms AFDM in BER over DD channels~\cite{ranasinghe2025afbm_receiver}, so securing it does not cost the legitimate user reliability. AFBM also contains two DAFT stages, one for frequency spreading and one inside the precoder. We therefore propose a double chirp-permutation, with one secret key per stage, and examine a single-permutation variant in which one key permutes the same chirp sequence in both stages. A wrong key is conjugated by a Fourier stage and becomes a dense matrix, so the precoder no longer restores complex orthogonality and the filter-bank interference leaks into the detector. In the double scheme, an eavesdropper with either key wrong has a BER between $0.47$ and $0.51$ over the whole SNR range and a rate below $0.09$~bit/symbol, even with perfect channel state information (CSI) and co-location with the legitimate receiver, and blind carrier-phase recovery does not change this. The price is a $1.9$~dB increase in peak-to-average power ratio (PAPR), and a search space of $10^{93.3}$ against a key space of $10^{96.5}$.

Section~\ref{sec:model} presents the AFBM system model and Section~\ref{sec:scheme} the proposed schemes. Section~\ref{sec:results} gives the numerical results and Section~\ref{sec:conclusion} concludes.

\section{System Model}
\label{sec:model}

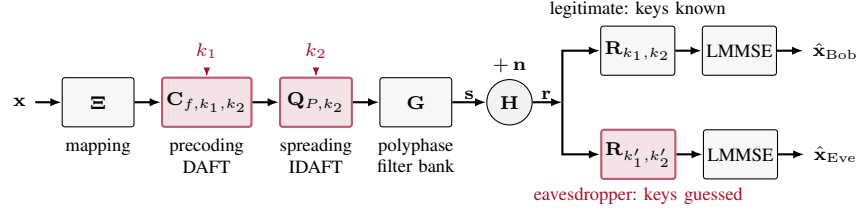
\begin{figure*}[t]
    \centering
    \begin{tikzpicture}[
  font=\scriptsize,
  blk/.style={draw, rounded corners=1pt, minimum height=6.5mm, minimum width=9.5mm,
              align=center, inner sep=1.5pt, fill=black!3},
  keyblk/.style={blk, fill=myred!12, draw=myred!70, thick},
  ch/.style={draw, circle, minimum size=6mm, inner sep=0pt, fill=black!5},
  ar/.style={-{Latex[length=1.5mm]}, thick},
  key/.style={myred, font=\scriptsize\bfseries},
  lbl/.style={font=\scriptsize, inner sep=1pt}
]

\node (x)   at (0,0) {$\mathbf{x}$};
\node[blk, right=3.5mm of x]      (xi)  {$\bm{\Xi}$};
\node[keyblk, right=3.5mm of xi]  (cf)  {$\mathbf{C}_{f,k_1,k_2}$};
\node[keyblk, right=3.5mm of cf]  (qp)  {$\mathbf{Q}_{P,k_2}$};
\node[blk, right=3.5mm of qp]     (g)   {$\mathbf{G}$};
\node[ch,  right=4.5mm of g]      (h)   {$\mathbf{H}$};

\draw[ar] (x)  -- (xi);
\draw[ar] (xi) -- (cf);
\draw[ar] (cf) -- (qp);
\draw[ar] (qp) -- (g);
\draw[ar] (g)  -- node[lbl, above]{$\mathbf{s}$} (h);

\node[key, above=1.2mm of cf] (kk1) {$k_1$};
\node[key, above=1.2mm of qp] (kk2) {$k_2$};
\draw[myred, -{Latex[length=1.2mm]}] (kk1) -- (cf);
\draw[myred, -{Latex[length=1.2mm]}] (kk2) -- (qp);

\node[lbl, below=1.0mm of xi, text width=13mm, align=center] {mapping};
\node[lbl, below=1.0mm of cf, text width=17mm, align=center] {precoding\\ DAFT};
\node[lbl, below=1.0mm of qp, text width=17mm, align=center] {spreading\\ IDAFT};
\node[lbl, below=1.0mm of g,  text width=14mm, align=center] {polyphase\\ filter bank};

\node[lbl, above=0.5mm of h] {$+\,\mathbf{n}$};

\coordinate (split) at ($(h)+(7mm,0)$);
\draw[ar] (h) -- node[lbl, above, pos=0.45]{$\mathbf{r}$} (split);

\node[blk, above right=3.5mm and 5mm of split] (rb) {$\mathbf{R}_{k_1,k_2}$};
\node[keyblk, below right=3.5mm and 5mm of split] (re) {$\mathbf{R}_{k'_1,k'_2}$};
\draw[ar] (split) |- (rb);
\draw[ar] (split) |- (re);

\node[blk, right=3.5mm of rb] (eb) {LMMSE};
\node[blk, right=3.5mm of re] (ee) {LMMSE};
\draw[ar] (rb) -- (eb);
\draw[ar] (re) -- (ee);

\node[right=3.5mm of eb] (xb) {$\hat{\mathbf{x}}_{\mathrm{Bob}}$};
\node[right=3.5mm of ee] (xe) {$\hat{\mathbf{x}}_{\mathrm{Eve}}$};
\draw[ar] (eb) -- (xb);
\draw[ar] (ee) -- (xe);

\node[lbl, above=0.6mm of rb, align=center] {legitimate: keys known};
\node[lbl, below=0.6mm of re, align=center, text=myred] {eavesdropper: keys guessed};

\end{tikzpicture}
    \caption{Double chirp-permuted AFBM transceiver. The secret key $k_1$ enters
    the precoding DAFT inside $\mathbf{C}_{f,k_1,k_2}$ and $k_2$ enters the
    spreading IDAFT $\mathbf{Q}_{P,k_2}$. Each key permutes the chirp sequence
    that its transform interposes between two Fourier stages, so that a mismatch
    in either one is spread into a dense product rather than localized on a
    diagonal. Bob holds the keys, Eve guesses them.}
    \label{fig:system_model}
\end{figure*}

Fig.~\ref{fig:system_model} depicts the transceiver. Consider a transmitter (Alice), a legitimate
receiver (Bob) and a passive eavesdropper (Eve), all with single antennas. Alice
has no \gls{CSI} of either link, which rules out \gls{CSI}-dependent
techniques such as artificial noise or secrecy beamforming.
We adopt a worst-case paradigm in which Bob and Eve both have
perfect \gls{CSI} of their channels. To probe the intrinsic
robustness of the scheme, Eve is also assumed to be co-located with
Bob, so that $\mathbf{H}_E=\mathbf{H}_B$ and she receives the identical signal~\cite{rou2025chirppermuted}. Alice and Bob
share the secret keys that govern the lexicographic permutations
applied to the \gls{DAFT} chirp sequences; Eve knows every other system
parameter but not the keys.

\subsection{Transmit Signal Model}

Let $L$ be the number of active subcarriers of an \gls{AFDM} system with $N$
total subcarriers, organized into $K$ blocks, and let
$\mathbf{x} \in \mathcal{D}^{KL/2 \times 1}$ collect the \gls{QAM} transmit
symbols. We summarize the \gls{AFBM} transmitter of~\cite{ranasinghe2026afbm,
senger2025afbm_waveform}, to which we refer for the full derivation, and keep
only what the security analysis needs.

The $L$-point \gls{DAFT} matrix is
\begin{equation}
\mathbf{W}_{L} = \mathbf{\Lambda}_{c_1,L}\mathbf{F}_{L}\mathbf{\Lambda}_{c_2,L},
\quad
\mathbf{\Lambda}_{c_i,L} = \mathrm{diag}\big[e^{-j2\pi c_i n^2}\big]_{n=0}^{L-1},
\label{eq:daft}
\end{equation}
with $\mathbf{F}_{L}$ the normalized \gls{DFT} matrix. Since $\mathbf{W}_L$ is
used as a \emph{spreading} operator in \eqref{eq:cf}, $\mathbf{\Lambda}_{c_2,L}$
multiplies the data vector directly, whereas $\mathbf{\Lambda}_{c_1,L}$ is the
factor placed between $\mathbf{F}_L$ and the subsequent spreading stage. This
position matters for the security analysis of Section~\ref{sec:whatprotects}. The chirp
rate $c_1$ is also constrained by the full-diversity condition
$c_1=(2(f_{\max}+\Gamma)+1)/2N$, which affects the key space (Section~\ref{sec:keyspace}).

Frequency containment is obtained by an \gls{IDAFT} whose output is zero-padded
in the frequency domain~\cite{savaux2024special},
\begin{equation}
\mathbf{Q}_{P} = \mathbf{F}_{N}^{H}\mathbf{T}_{NP}\mathbf{F}_P \mathbf{\tilde{W}}\herm_{P} \in \mathbb{C}^{N\times L},
\quad
\mathbf{\tilde{W}}_{P} = [\mathbf{I}_L \; \mathbf{0}]\,\mathbf{W}_P,
\label{eq:Q_P}
\end{equation}
where $\mathbf{T}_{NP}$ is the $N\times P$ zero-padding selection matrix and
$L<P<N$, so that the precoding \gls{DAFT} does not cancel the \gls{IDAFT} of the
filter bank. We write $P_{\mathrm{ch}}$ for the number of channel paths to avoid
overloading $P$.

Per-subcarrier filtering is applied through a \gls{PPN} built from a prototype
filter $\mathbf{g}$ of length $ON$, whose polyphase blocks
$\mathbf{G}_p=\mathrm{diag}(\mathbf{g}_p)$ assemble the single-symbol matrix
$\mathbf{\widetilde{G}}\in\mathbb{R}^{ON\times N}$ and the block-Toeplitz frame
matrix $\mathbf{G}$, in which consecutive symbols are delayed by $N/2$
samples~\cite{pereira2022generalized}. Because standard \gls{FBMC} lacks complex
orthogonality, a precoder $\mathbf{C}_f$ is required, following the
DFT-precoded filter bank construction
of~\cite{nissel2018pruned}, to satisfy
\begin{equation}
\mathbf{C}\herm_f\mathbf{Q}_{P}^{H}\mathbf{\tilde{G}}^{H}\mathbf{\tilde{G}}\mathbf{Q}_{P}\mathbf{C}_f \approx \mathbf{U},
\label{eq:complex_orthogonality}
\end{equation}
with $\mathbf{U}$ diagonal and unitary on $L/2$ positions. Taking
\begin{equation}
\mathbf{C}_f \triangleq \mathbf{W}_{L} \mathrm{diag}\{\tilde{\mathbf{b}}\},
\qquad
[\mathbf{\tilde{b}}]_{\tilde{l}} = [\mathbf{\tilde{c}}]_{\tilde{l}}^{-1/2}
\label{eq:cf}
\end{equation}
on the first and last $L/4$ positions and zero elsewhere, with
$\mathbf{\tilde{c}} = \mathrm{diag}\{\mathbf{W}\herm_L\mathbf{Q}_{P}^{H}\mathbf{\widetilde{G}}^H\mathbf{\widetilde{G}}\mathbf{Q}_{P}\mathbf{W}_L\}$,
forces the diagonal of \eqref{eq:complex_orthogonality} to unity. Correct
compensation requires the intrinsic interference to be confined to a single
coefficient, which caps the overlap factor at $O\le1.5$~\cite{pereira2022generalized};
consecutive blocks are then transmitted every $N/2$ samples~\cite{Ranasinghe_ICNC2025_oversampling}.

Collecting the mapping $\bm{\Xi}=\mathbf{I}_K\otimes\bar{\bm{\Xi}}$ that places
the symbols on the active positions, the transmit signal for all $K$ blocks is
\begin{equation}
\label{eq:td_tx_signal}
\mathbf{s} = \mathbf{G} \big(\mathbf{I}_{K} \otimes \mathbf{Q}_{P} \mathbf{C}_f \big) \bm{\Xi} \mathbf{x} \in \mathbb{C}^{M \times 1},
\end{equation}
with $M = ON + \tfrac{N}{2}(K-1)$. A \gls{CP} is not required, since the
prototype filter is well localized in time.

\subsection{Channel and Receiver Model}
\label{sec:chanmodel}

The signal propagates over a \gls{DD} channel with $P_{\mathrm{ch}}$ paths. With
a prefix of length at least $\ell_{\max}$ appended to the frame, the discrete channel is the $M\times M$ circular matrix
\begin{equation}
\mathbf{H} = \sum_{p=1}^{P_{\mathrm{ch}}} h_p\, \bm{\Gamma}_{p}\, \bm{\Delta}_{f_p}\, \bm{\Pi}^{\ell_p},
\label{eq:channel}
\end{equation}
where $\bm{\Pi}$ is the cyclic delay matrix,
$\bm{\Delta}_{f_p}=\mathrm{diag}\big(e^{-j2\pi f_p n/N}\big)_{n=0}^{M-1}$ is the
fractional-Doppler matrix, and $\bm{\Gamma}_p$ is the prefix phase compensation,
equal to $\mathbf{I}$ for a plain \gls{CP} and to
$\mathrm{diag}\big([e^{-j2\pi c_1(M^2-2Mq)}]_{q=\ell_p}^{1},\mathbf{1}_{M-\ell_p}\big)$
for a \gls{CPP}. The gains satisfy $h_p\sim\mathcal{CN}(0,1/P_{\mathrm{ch}})$,
the delays $\ell_p$ are uniform on $[0,\ell_{\max}]$ and the Doppler shifts
$f_p$ are uniform on $[-f_{\max},f_{\max}]$.

The received signal is $\mathbf{r} = \mathbf{H}\mathbf{s} + \mathbf{n}$, with
$\mathbf{s}$ as in \eqref{eq:td_tx_signal} and
$\mathbf{n}\sim\mathcal{CN}(\mathbf{0},\sigma_n^2\mathbf{I}_M)$. Writing the
complete transmit and matched receive operators as
\begin{equation}
\mathbf{T}_{k_1,k_2} = \mathbf{G}\big(\mathbf{I}_{K}\otimes\mathbf{Q}_{P,k_2}\mathbf{C}_{f,k_1,k_2}\big)\bm{\Xi},
\qquad
\mathbf{R}_{k_1,k_2} = \mathbf{T}_{k_1,k_2}\herm,
\label{eq:TR}
\end{equation}
the detector input is $\mathbf{y}=\mathbf{R}\mathbf{r}$ with effective channel
$\mathbf{H}_{\mathrm{eff}}=\mathbf{R}\mathbf{H}\mathbf{T}$, and the
\gls{LMMSE} detector is
\begin{equation}
\mathbf{x}_{\mathrm{est}} =
\big(\mathbf{H}_{\mathrm{eff}}\herm\mathbf{H}_{\mathrm{eff}}+\sigma_n^2\mathbf{I}\big)^{-1}\mathbf{H}_{\mathrm{eff}}\herm\mathbf{y},
\label{eq:lmmse}
\end{equation}
which treats the noise at the detector input as white. The simulations use the
exact covariance $\sigma_n^2\mathbf{R}\mathbf{R}\herm$ of the coloured noise.

\section{Proposed Physical Layer Security Schemes}
\label{sec:scheme}

\subsection{Double Chirp-Permutation in AFBM}

Let Alice and Bob share two secret permutation keys,
$k_1 \in \{1,\dots,L!\}$ and $k_2 \in \{1,\dots,P!\}$, and let
$\mathrm{perm}(\bm{\lambda}, k)$ denote the $k$-th lexicographic permutation of
$\bm{\lambda}$, following the construction of
CP-\gls{AFDM}~\cite{rou2025chirppermuted, rou2025cpafdm_versatile}.
Each key is applied to the chirp sequence that the \gls{DAFT} of \eqref{eq:daft}
interposes between its \gls{DFT} and the following operator; the cost of this
placement is quantified in Section~\ref{sec:res_papr}. For the precoding \gls{DAFT},
\begin{equation}
\mathbf{\Lambda}_{c_1,L,k_1} \triangleq \mathrm{diag}(\mathrm{perm}(\bm{\lambda}_{c_1,L}, k_1)) \in \mathbb{C}^{L \times L},
\label{eq:permL}
\end{equation}
yielding
$\mathbf{W}_{L,k_1} = \mathbf{\Lambda}_{c_1,L,k_1}\mathbf{F}_{L}\mathbf{\Lambda}_{c_2,L}$,
which enters the precoder through \eqref{eq:cf}. For the spreading \gls{DAFT},
\begin{equation}
\mathbf{\Lambda}_{c_2,P,k_2} \triangleq \mathrm{diag}(\mathrm{perm}(\bm{\lambda}_{c_2,P}, k_2)) \in \mathbb{C}^{P \times P},
\label{eq:permP}
\end{equation}
yielding $\mathbf{W}_{P,k_2}$ and, through \eqref{eq:Q_P}, the secured
$\mathbf{Q}_{P,k_2}$. In \eqref{eq:permL} the permuted diagonal is placed
between $\mathbf{F}_L$ and $\mathbf{Q}_P$, and in \eqref{eq:permP} between
$\mathbf{F}_P$ and $\mathbf{F}_P\herm$. Section~\ref{sec:whatprotects} shows
that this common placement gives both keys their effect.

The compensation vector is defined from the whole structure and must in general
be recomputed from both keys,
$\mathbf{\tilde{c}}_{k_1,k_2} = \mathrm{diag}\{\mathbf{W}\herm_{L,k_1}\mathbf{Q}_{P,k_2}^{H}\mathbf{\widetilde{G}}^H\mathbf{\widetilde{G}}\mathbf{Q}_{P,k_2}\mathbf{W}_{L,k_1}\}$,
giving $\mathbf{C}_{f,k_1,k_2} \triangleq \mathbf{W}_{L,k_1} \mathrm{diag}\{\tilde{\mathbf{b}}_{k_1,k_2}\}$.
For a power-complementary prototype this simplifies. Every
polyphase block is diagonal, so $\mathbf{\widetilde{G}}\herm\mathbf{\widetilde{G}}
= \mathrm{diag}(\mathbf{d})$ with $d_n=\sum_k |g[n+kN]|^2$, and $\mathbf{d}$ is
constant. Together with $\mathbf{Q}_{P,k_2}\herm\mathbf{Q}_{P,k_2}=\mathbf{I}_L$ and the unitarity of
$\mathbf{W}_{L,k_1}$, this gives $\mathbf{\tilde{c}}=d\,\mathbf{1}$, independent of both keys.
The security therefore comes from $\mathbf{W}_{L,k_1}$ and
$\mathbf{Q}_{P,k_2}$ themselves, as confirmed numerically in Section~\ref{sec:res_secrecy}.

\subsection{Single Chirp-Permutation in AFBM}
\label{sec:single}

A simpler variant uses one key $k\in\{1,\dots,P!\}$ and permutes only the $c_2$ chirp. The sequence
$\bm{\lambda}_{c_2,L}$ is the first $L$ entries of $\bm{\lambda}_{c_2,P}$, so the two
DAFTs share the same $c_2$ chirp and one key can serve both. The key $k$ permutes
$\bm{\lambda}_{c_2,P}$ in $\mathbf{W}_{P}$, and its restriction $\bar{k}$ permutes $\bm{\lambda}_{c_2,L}$ in
$\mathbf{W}_{L}$, where $\bar{k}$ is the permutation of $\{0,\dots,L-1\}$ obtained by
deleting from the $k$-th permutation the entries not smaller than $L$:
\begin{equation}
\mathbf{W}_{L,k} = \mathbf{\Lambda}_{c_1,L}\mathbf{F}_L\mathbf{\Lambda}_{c_1,L,\bar{k}},
\qquad
\mathbf{W}_{P,k} = \mathbf{\Lambda}_{c_1,P}\mathbf{F}_P\mathbf{\Lambda}_{c_1,P,k}.
\label{eq:single}
\end{equation}
The permuted chirp sits in a different position in each stage. In $\mathbf{W}_{L,k}$ it multiplies the
data vector, so a mismatch is a unit-modulus diagonal on the data, which commutes out of the product
as a per-subcarrier phase rotation. In $\mathbf{W}_{P,k}$ it lies between $\mathbf{F}_P$ and
$\mathbf{F}_P\herm$ once \eqref{eq:Q_P} is formed, as in the double scheme. The protection of the single
permutation therefore comes from the spreading stage alone. The scheme needs one key instead of two and leaves
the PAPR of $\mathbf{W}_{L}$ untouched, because $\bm{\lambda}_{c_1,L}$ is not permuted.

\subsection{Mismatched Effective Channel}
\label{sec:whatprotects}

With $\mathbf{T}_{k_1,k_2}$ and $\mathbf{R}_{k_1,k_2}=\mathbf{T}\herm_{k_1,k_2}$
as in \eqref{eq:TR}, Bob faces
$\mathbf{G}_{\mathrm{Bob}}=\mathbf{R}_{k_1,k_2}\mathbf{H}\mathbf{T}_{k_1,k_2}$
and an eavesdropper guessing $(k'_1,k'_2)$ faces
$\mathbf{G}_{\mathrm{Eve}}=\mathbf{R}_{k'_1,k'_2}\mathbf{H}\mathbf{T}_{k_1,k_2}$.
The intrinsic interference between overlapping blocks is not conditioned by \eqref{eq:cf}, which uses the
single-symbol matrix $\mathbf{\widetilde{G}}$ alone. It limits AFBM at high \gls{SNR} and is common to Bob and Eve.

Write $\mathbf{M}=\mathbf{Q}_{P}\herm\mathbf{\widetilde{G}}\herm\mathbf{\widetilde{G}}\mathbf{Q}_{P}\in\mathbb{C}^{L\times L}$
for the Gram matrix that \eqref{eq:complex_orthogonality} conditions, and
consider a mismatch in $k_1$ alone. With $\mathbf{A}=\mathbf{\Lambda}_{c_2,L}\mathrm{diag}\{\tilde{\mathbf{b}}\}$
and $\mathbf{B}=\mathbf{\Lambda}\herm_{c_1,L,k_1}\mathbf{M}\mathbf{\Lambda}_{c_1,L,k_1}$, the product the
eavesdropper forms is $\mathbf{A}\herm\mathbf{F}\herm_L\mathbf{D}\mathbf{B}\,\mathbf{F}_L\mathbf{A}$, where
\begin{equation}
[\mathbf{D}]_{ll} = e^{-j2\pi c_1\left(\pi_{k_1}(l)^2-\pi_{k'_1}(l)^2\right)}
\label{eq:Dmismatch}
\end{equation}
is diagonal with unit modulus. Bob's own product is the case $\mathbf{D}=\mathbf{I}$, which \eqref{eq:cf}
conditions to the diagonal $\mathbf{U}$, so the eavesdropper's product splits into Bob's and an error term,
\begin{equation}
\mathbf{C}\herm_{f,k'_1}\mathbf{M}\,\mathbf{C}_{f,k_1} = \mathbf{U}+\mathbf{E},
\;\;
\mathbf{E} = \mathbf{A}\herm\mathbf{F}\herm_L(\mathbf{D}-\mathbf{I})\mathbf{B}\,\mathbf{F}_L\mathbf{A}.
\label{eq:UplusE}
\end{equation}
The mismatch enters $\mathbf{E}$ as a diagonal interposed between
$\mathbf{F}\herm_L$ and $\mathbf{B}$, so it cannot be commuted out to the edges of the
product. Conjugating a diagonal by $\mathbf{F}_L$ yields a circulant matrix, which is
diagonal only when the diagonal is constant. Since $\mathbf{D}-\mathbf{I}$
vanishes for $k'_1=k_1$ and is a non-constant unit-modulus diagonal
otherwise, a $k_1$ mismatch produces \gls{ICI} spread over all
subcarriers, and no per-subcarrier correction undoes it. For the parameters of Section~\ref{sec:results},
averaged over random guesses $k'_1$, $\|\mathbf{E}\|_F$ is $1.2$
times $\|\mathbf{U}\|_F$, with about a third of its energy off the diagonal.

The same structure governs $k_2$. From \eqref{eq:Q_P},
$\mathbf{\Lambda}\herm_{c_2,P,k_2}$ lies between $\mathbf{F}_P$ and
$\mathbf{F}_P\herm$, so a mismatch yields
$\mathbf{Q}\herm_{P,k'_2}\mathbf{\widetilde{G}}\herm\mathbf{\widetilde{G}}\mathbf{Q}_{P,k_2}$,
again a diagonal conjugated by a Fourier stage and hence dense. In both cases the precoding
compensation fails and the intrinsic interference of the filter bank leaks into the detector.

\subsection{Key Space versus Search Space}
\label{sec:keyspace}

The \emph{key space} counts the configurations Alice and Bob may agree upon, namely $L!\times P!$ for the
double scheme and $P!$ for the single one. The \emph{search space} counts the hypotheses Eve must test, and it
is smaller when distinct keys give the same transmit structure. Permutations that exchange equal entries of a
chirp sequence give the same diagonal, so the domain of $k_1$ contributes the multinomial
$L!/\prod_i m_i!$ over the multiplicities $m_i$ of $\bm{\lambda}_{c_1,L}$. Repeated entries are unavoidable:
the full-diversity condition makes $c_1$ rational, which makes $e^{-j2\pi c_1 n^2}$ periodic in $n^2$, and for
$L=32$ only $23$ of the $32$ entries are distinct. The chirp rate $c_2=1/(2\pi N)$ is irrational, so the
$P$ entries of $\bm{\lambda}_{c_2,P}$ are pairwise distinct. The domain of $k_2$ and the key of the single
permutation are therefore not reduced.

\subsection{Eavesdropper Model and Security Metrics}
\label{sec:metrics}

Besides BER, we use two measures. The first is an eavesdropper that applies carrier recovery before detection,
so that a degradation which is only a phase rotation would be removed. The blind fourth-power (Viterbi--Viterbi)
estimator~\cite{viterbi1983nonlinear} exploits $\Exp\{x^4\}=-1$ for \gls{QPSK} to form, per data subcarrier $j$,
\begin{equation}
\hat{\theta}_j = \tfrac{1}{4}\arg\Big(-\tfrac{1}{TK}\textstyle\sum_{t,k} \hat{x}_{t,k,j}^{\,4}\Big),
\label{eq:vv}
\end{equation}
where $\hat{x}_{t,k,j}$ are the detected symbols over $T$ frames and $K$ blocks, followed by a short known preamble
that resolves the residual $\pi/2$ ambiguity. It costs $\mathcal{O}(LT)$ and searches no keys.

The second is an achievable rate, since a BER close to $0.5$ constrains only hard decisions. Let $\gamma_i$ be the
post-equalization \gls{SINR} of stream $i$, obtained from the true response
$\mathbf{R}_{k'_1,k'_2}\mathbf{H}\mathbf{T}_{k_1,k_2}$ that the input contains while equalizing with the matrix
$\mathbf{R}_{k'_1,k'_2}\mathbf{H}\mathbf{T}_{k'_1,k'_2}$ that the eavesdropper can build. Then
\begin{equation}
C = \tfrac{1}{n}\textstyle\sum_i \log_2(1+\gamma_i),
\quad
R_s = \max\!\big(0, C_{\mathrm{Bob}} - C_{\mathrm{Eve}}\big)
\label{eq:secrecy}
\end{equation}
is the Gaussian-input rate achievable by an \gls{LMMSE} receiver. $R_s$ is \emph{not} the Wyner secrecy
capacity: each $C$ is a lower bound on the corresponding mutual information, and a difference of two lower bounds
bounds neither side of the difference.

\section{Numerical Results}
\label{sec:results}

\subsection{Setup}

We use $N=64$ subcarriers with $L=32$ active, an extended \gls{IDAFT} of size $P=48$, which meets the
$L<P<N$ requirement of \eqref{eq:Q_P}, and $K=8$ blocks of 4-\gls{QAM} symbols. The prototype is
power-complementary with overlap factor $O=1.5$. The channel has $P_{\mathrm{ch}}=3$ paths, a maximum delay
$\ell_{\max}=6$, a maximum Doppler shift $f_{\max}=3$ and an \gls{AFDM} guard band $\Gamma=1$, so that
$c_1=(2(f_{\max}+\Gamma)+1)/2N$ is rational and $c_2=1/(2\pi N)$ is irrational. The \gls{SNR} runs from $0$ to $21$~dB
in $3$~dB steps, with $2{,}000$ Monte Carlo trials for the BER and $300$ for the rate. The small $N$ keeps the
effective channel matrices exactly invertible. In each trial every waveform sees the same channel realization of
Section~\ref{sec:chanmodel}, so the comparison is paired. Detection is \gls{LMMSE}, and the
\gls{AFBM} transmit matrix is scaled so that $\Exp\{|s[m]|^2\}=1$, which makes $\sigma_n^2=1/\mathrm{SNR}$
the same $E_s/N_0$ for all waveforms.

\subsection{Security}
\label{sec:res_security}

\begin{figure}[!tb]
    \centering
    \setlength{\figurewidth}{0.84\columnwidth}%
    \setlength{\figureheight}{4.5cm}%
    \begin{tikzpicture}
\begin{axis}[
  width=\figurewidth, height=\figureheight, scale only axis,
  tick label style={font=\scriptsize},
  label style={font=\footnotesize},
  legend style={font=\scriptsize, draw=black!40, fill=white, inner sep=1.5pt,
                nodes={inner sep=1pt}, row sep=-0.5pt},
  legend cell align=left,
  grid=major, grid style={gray!25, line width=0.3pt},
  axis line style={black!80},
  every axis plot/.append style={line width=0.9pt},
  xlabel={SNR (dB)}, ylabel={BER},
  ymode=log, ymin=1e-3, ymax=1, xmin=0, xmax=21, xtick={0,3,...,21},
  legend pos=south west
]
\addplot[color=black, solid, mark=o, mark size=1.6pt, mark options={solid, fill=white}] coordinates {(0,0.198314) (3,0.14265) (6,0.0970957) (9,0.0585996) (12,0.0312695) (15,0.0140215) (18,0.00556641) (21,0.00200391)};
\addlegendentry{Bob}
\addplot[color=myblue, solid, mark=square, mark size=1.6pt, mark options={solid, fill=white}] coordinates {(0,0.504664) (3,0.503619) (6,0.505074) (9,0.505426) (12,0.504826) (15,0.505914) (18,0.505527) (21,0.504066)};
\addlegendentry{Eve, both keys wrong}
\addplot[color=myyellow, solid, mark=triangle, mark size=1.6pt, mark options={solid, fill=white}] coordinates {(0,0.474537) (3,0.472936) (6,0.472652) (9,0.472359) (12,0.474027) (15,0.4765) (18,0.477705) (21,0.481764)};
\addlegendentry{Eve, $k_2$ wrong}
\addplot[color=myred, solid, mark=diamond, mark size=1.6pt, mark options={solid, fill=white}] coordinates {(0,0.487738) (3,0.487525) (6,0.487568) (9,0.487449) (12,0.489316) (15,0.48952) (18,0.490982) (21,0.491127)};
\addlegendentry{Eve, $k_1$ wrong}
\addplot[color=mygreen, dashed, mark=pentagon, mark size=1.6pt, mark options={solid, fill=white}] coordinates {(0,0.48784) (3,0.476842) (6,0.491018) (9,0.474809) (12,0.474016) (15,0.480236) (18,0.486482) (21,0.488457)};
\addlegendentry{Eve, $k_1$ wrong, attack}
\addplot[color=mypurple, dashed, mark=asterisk, mark size=1.6pt, mark options={solid, fill=white}] coordinates {(0,0.483832) (3,0.473359) (6,0.47775) (9,0.480121) (12,0.47667) (15,0.483111) (18,0.487979) (21,0.486293)};
\addlegendentry{Eve, both wrong, attack}
\end{axis}
\end{tikzpicture}
    \caption{Double CP-AFBM: legitimate receiver versus key-mismatched
    eavesdroppers, with and without the blind attack of \eqref{eq:vv}. Every
    mismatched configuration stays near $0.5$, and the attack does not move it.}
    \label{fig:ber_pls}
\end{figure}
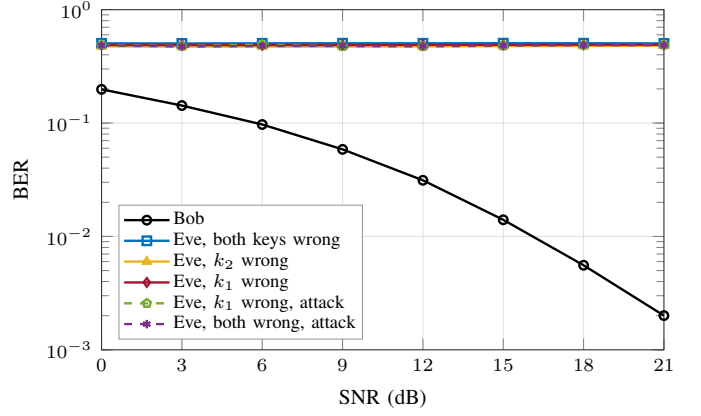

Fig.~\ref{fig:ber_pls} shows the \gls{BER} of Bob and of eavesdroppers with wrong keys. Every mismatched
configuration stays between $0.47$ and $0.51$ over the whole \gls{SNR} range, while Bob's BER falls from $0.20$ to
$2.0\times10^{-3}$. The same holds for each key alone: a mismatch confined to $k_1$ leaves Eve between $0.48$ and
$0.50$, and one confined to $k_2$ between $0.47$ and $0.49$. This occurs although Eve has perfect \gls{CSI}, is
co-located with Bob and knows every parameter except the keys. The blind attack of
Section~\ref{sec:metrics} does not help, since the attacked curves remain between $0.47$ and $0.50$. The
degradation is therefore the circulant interference of Section~\ref{sec:whatprotects} and not a rotation that carrier
recovery could remove. A \gls{BER} at this level is far above the threshold of any practical wiretap
code~\cite{klinc2011ldpc}.

For the single permutation, simulated with $1{,}000$ trials and not shown in Fig.~\ref{fig:ber_pls}, a wrong key leaves Eve
between $0.46$ and $0.47$, while Bob's BER falls from $0.17$ to $2.8\times10^{-3}$. The blind attack, however,
lowers Eve to between $0.27$ and $0.33$. Part of the mismatch in $\mathbf{W}_{L,k}$ is a per-subcarrier phase rotation
(Section~\ref{sec:single}), and the estimator removes it. Eve is still far from a usable error rate, but the single
permutation is weaker than the double scheme against this attack.

\subsection{Secrecy Rate}
\label{sec:res_secrecy}

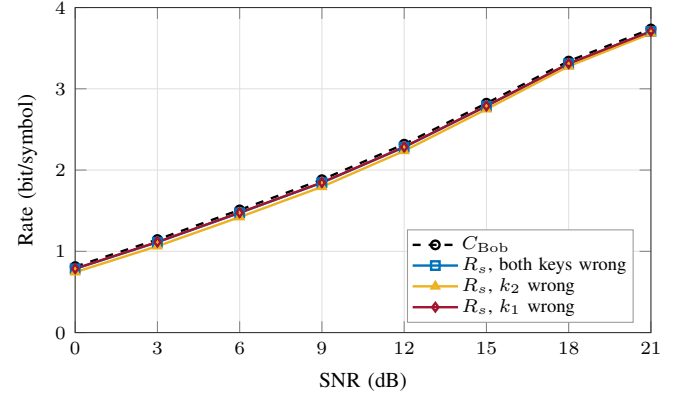
\begin{figure}[!tb]
    \centering
    \setlength{\figurewidth}{0.86\columnwidth}%
    \setlength{\figureheight}{4.3cm}%
    \begin{tikzpicture}
\begin{axis}[
  width=\figurewidth, height=\figureheight, scale only axis,
  tick label style={font=\scriptsize},
  label style={font=\footnotesize},
  legend style={font=\scriptsize, draw=black!40, fill=white, inner sep=1.5pt,
                nodes={inner sep=1pt}, row sep=-0.5pt},
  legend cell align=left,
  grid=major, grid style={gray!25, line width=0.3pt},
  axis line style={black!80},
  every axis plot/.append style={line width=0.9pt},
  xlabel={SNR (dB)}, ylabel={Rate (bit/symbol)},
  xmin=0, xmax=21, xtick={0,3,...,21}, ymin=0, ymax=4,
  legend pos=south east
]
\addplot[color=black, dashed, mark=o, mark size=1.6pt, mark options={solid, fill=white}] coordinates {(0,0.811368) (3,1.14493) (6,1.5096) (9,1.88147) (12,2.31838) (15,2.8222) (18,3.34083) (21,3.73589)};
\addlegendentry{$C_{\mathrm{Bob}}$}
\addplot[color=myblue, solid, mark=square, mark size=1.6pt, mark options={solid, fill=white}] coordinates {(0,0.784446) (3,1.11185) (6,1.47313) (9,1.84395) (12,2.2814) (15,2.78662) (18,3.30793) (21,3.70608)};
\addlegendentry{$R_s$, both keys wrong}
\addplot[color=myyellow, solid, mark=triangle, mark size=1.6pt, mark options={solid, fill=white}] coordinates {(0,0.742961) (3,1.06347) (6,1.42122) (9,1.79377) (12,2.23604) (15,2.7489) (18,3.27768) (21,3.68189)};
\addlegendentry{$R_s$, $k_2$ wrong}
\addplot[color=myred, solid, mark=diamond, mark size=1.6pt, mark options={solid, fill=white}] coordinates {(0,0.787032) (3,1.11506) (6,1.47615) (9,1.84693) (12,2.28436) (15,2.78929) (18,3.31009) (21,3.70775)};
\addlegendentry{$R_s$, $k_1$ wrong}
\end{axis}
\end{tikzpicture}
    \caption{Secrecy rate for the three key-mismatch cases, with Bob's
    achievable rate as reference. The eavesdropper's rate, which is the gap
    between each $R_s$ curve and $C_{\rm Bob}$, stays below $0.09$~bit/symbol
    and is essentially flat in SNR, so $R_s$ tracks $C_{\rm Bob}$ almost
    exactly.}
    \label{fig:secrecy}
\end{figure}

Fig.~\ref{fig:secrecy} shows the rates defined in Section~\ref{sec:metrics}. Eve's rate stays below
$0.09$~bit/symbol over the whole range, from $8.4\%$ of Bob's rate at $0$~dB to $1.4\%$ at $21$~dB, and it is nearly
flat in \gls{SNR}, peaking near $6$ to $9$~dB. Raising the transmit power does not help Eve, because the residual
filter-bank interference scales with the signal and not with the noise. The secrecy rate therefore follows Bob's
rate, from $0.74$~bit/symbol at $0$~dB to $3.68$~bit/symbol at $21$~dB in the least favourable of the three
mismatch cases, and to $3.71$ in the most favourable.

We also check the property stated in Section~\ref{sec:scheme}. With the power-complementary prototype the relative
spread of $\mathbf{\tilde{c}}_{k_1,k_2}$ over the active positions is of order $10^{-16}$, so $\tilde{\mathbf{b}}$
is identical for all key pairs. With a PHYDYAS prototype, whose polyphase energy is not flat, the spread is
$1.1\times10^{-1}$ and $\tilde{\mathbf{b}}$ varies with both keys.

\subsection{Exhaustive Search Complexity}
\label{sec:res_complexity}

\begin{figure}[!tb]
    \centering
    \setlength{\figurewidth}{0.74\columnwidth}%
    \setlength{\figureheight}{5.2cm}%
    \begin{tikzpicture}
\begin{axis}[
  width=\figurewidth, height=\figureheight, scale only axis,
  tick label style={font=\scriptsize},
  label style={font=\footnotesize},
  legend style={font=\scriptsize, draw=black!40, fill=white, inner sep=1.5pt,
                nodes={inner sep=1pt}, row sep=-0.5pt},
  legend cell align=left,
  grid=major, grid style={gray!25, line width=0.3pt},
  axis line style={black!80},
  every axis plot/.append style={line width=0.9pt},
  xlabel={Number of active subcarriers $L$},
  ylabel={$\log_{10}$(hypotheses or queries)},
  xmin=4, xmax=64, ymin=-5, ymax=250, clip=false,
  legend columns=3, legend style={at={(0.5,1.03)}, anchor=south,
  /tikz/every even column/.append style={column sep=2.5mm}}
]
\addplot[color=black, solid, forget plot] coordinates {(4,1.20412) (6,1.5563) (8,1.80618) (10,2) (12,2.15836) (14,2.29226) (16,2.40824) (18,2.51055) (20,2.60206) (22,2.68485) (24,2.76042) (26,2.82995) (28,2.89432) (30,2.95424) (32,3.0103) (34,3.06296) (36,3.11261) (38,3.15957) (40,3.20412) (42,3.2465) (44,3.28691) (46,3.32552) (48,3.36248) (50,3.39794) (52,3.43201) (54,3.46479) (56,3.49638) (58,3.52686) (60,3.5563) (62,3.58478) (64,3.61236)};
\addplot[color=myblue, solid, forget plot] coordinates {(4,1.38021) (6,2.85733) (8,4.60552) (10,6.55976) (12,8.68034) (14,10.9404) (16,13.3206) (18,15.8063) (20,18.3861) (22,21.0508) (24,23.7927) (26,26.6056) (28,29.4841) (30,32.4237) (32,35.4202) (34,38.4702) (36,41.5705) (38,44.7185) (40,47.9116) (42,51.1477) (44,54.4246) (46,57.7406) (48,61.0939) (50,64.4831) (52,67.9066) (54,71.3633) (56,74.8519) (58,78.3712) (60,81.9202) (62,85.4979) (64,89.1034)};
\addplot[color=myblue, dashed, forget plot] coordinates {(4,0.690105) (6,1.42867) (8,2.30276) (10,3.27988) (12,4.34017) (14,5.4702) (16,6.6603) (18,7.90315) (20,9.19305) (22,10.5254) (24,11.8963) (26,13.3028) (28,14.7421) (30,16.2118) (32,17.7101) (34,19.2351) (36,20.7853) (38,22.3592) (40,23.9558) (42,25.5739) (44,27.2123) (46,28.8703) (48,30.5469) (50,32.2415) (52,33.9533) (54,35.6816) (56,37.426) (58,39.1856) (60,40.9601) (62,42.749) (64,44.5517)};
\addplot[color=myorange, solid, forget plot] coordinates {(4,2.85733) (6,5.55976) (8,8.68034) (10,12.1165) (12,15.8063) (14,19.7083) (16,23.7927) (18,28.037) (20,32.4237) (22,36.9387) (24,41.5705) (26,46.3096) (28,51.1477) (30,56.0778) (32,61.0939) (34,66.1906) (36,71.3633) (38,76.6077) (40,81.9202) (42,87.2972) (44,92.7359) (46,98.2333) (48,103.787) (50,109.395) (52,115.054) (54,120.763) (56,126.52) (58,132.324) (60,138.172) (62,144.063) (64,149.996)};
\addplot[color=myorange, dashed, forget plot] coordinates {(4,1.42867) (6,2.77988) (8,4.34017) (10,6.05825) (12,7.90315) (14,9.85415) (16,11.8963) (18,14.0185) (20,16.2118) (22,18.4693) (24,20.7853) (26,23.1548) (28,25.5739) (30,28.0389) (32,30.5469) (34,33.0953) (36,35.6816) (38,38.3038) (40,40.9601) (42,43.6486) (44,46.368) (46,49.1166) (48,51.8935) (50,54.6975) (52,57.527) (54,60.3815) (56,63.26) (58,66.162) (60,69.086) (62,72.0315) (64,74.998)};
\addplot[color=myred, solid, forget plot] coordinates {(4,4.23754) (6,7.33791) (8,12.9848) (10,18.0742) (12,21.3741) (14,29.7457) (16,35.9092) (18,34.8242) (20,48.4015) (22,56.4843) (24,57.837) (26,71.109) (28,77.0195) (30,78.595) (32,93.3277) (34,102.253) (36,90.6948) (38,118.617) (40,123.158) (42,124.125) (44,141.14) (46,152.663) (48,147.226) (50,167.709) (52,175.736) (54,156.182) (56,191.512) (58,206.481) (60,194.121) (62,225.046) (64,231.171)};
\addplot[color=myred, dashed, forget plot] coordinates {(4,2.11877) (6,3.66895) (8,6.4924) (10,9.0371) (12,10.687) (14,14.8728) (16,17.9546) (18,17.4121) (20,24.2007) (22,28.2421) (24,28.9185) (26,35.5545) (28,38.5097) (30,39.2975) (32,46.6638) (34,51.1265) (36,45.3474) (38,59.3085) (40,61.579) (42,62.0625) (44,70.57) (46,76.3315) (48,73.613) (50,83.8545) (52,87.868) (54,78.091) (56,95.756) (58,103.24) (60,97.0605) (62,112.523) (64,115.585)};
\addplot[gray, dotted, line width=0.8pt, forget plot] coordinates {(4,15) (64,15)};
\addplot[gray, dotted, line width=0.8pt, forget plot] coordinates {(4,30) (64,30)};
\addplot[gray!70, densely dotted, line width=0.5pt, forget plot] coordinates {(32,-5) (32,250)};
\node[font=\scriptsize, text=gray!50!black, anchor=west, inner sep=1pt] at (axis cs:64,13) {$10^{15}$};
\node[font=\scriptsize, text=gray!50!black, anchor=west, inner sep=1pt] at (axis cs:64,32) {$10^{30}$};
\node[font=\scriptsize, text=gray!50!black, anchor=north west, inner sep=1.5pt, rotate=90] at (axis cs:32.3,165) {$L=32$};
\addlegendimage{black, solid}\addlegendentry{AFDM}
\addlegendimage{myblue, solid}\addlegendentry{CP-AFDM}
\addlegendimage{myorange, solid}\addlegendentry{single CP-AFBM}
\addlegendimage{myred, solid}\addlegendentry{double CP-AFBM}
\addlegendimage{gray, solid}\addlegendentry{classical search}
\addlegendimage{gray, dashed}\addlegendentry{Grover search}
\end{axis}
\end{tikzpicture}
    \caption{Brute-force attack complexity versus system size, $P=3L/2$. The
    key space and the search space of the double CP-AFBM scheme and the key
    space of the single permutation are reported, together with the
    corresponding Grover query counts.}
    \label{fig:computational_growth}
\end{figure}
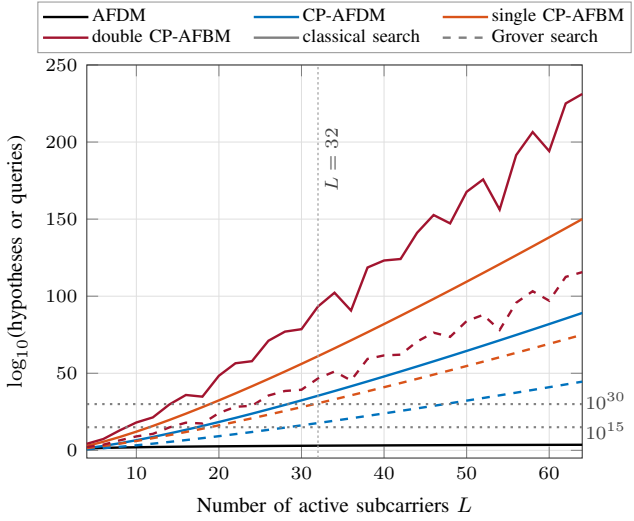

Fig.~\ref{fig:computational_growth} shows the key space and the search space against $L$, with $P=3L/2$ as implied
by $L=N/2$ and $P=3N/4$. At $L=32$ the double scheme has a key space of $10^{96.5}$ and a search space of
$10^{93.3}$, the $3.2$-order gap being the nine repeated entries of $\bm{\lambda}_{c_1,L}$. The single permutation has
a key space and a search space of $P!=10^{61.1}$, since $c_2$ is irrational. A quantum adversary using \gls{GAS}~\cite{gilliam2021grover}
still needs $\mathcal{O}(\sqrt{\cdot})$ queries~\cite{choi2023quantum}: $10^{46.7}$ over the double search space,
$10^{48.3}$ over the double key space and $10^{30.5}$ for the single permutation. At an assumed rate of
$10^{9}$ queries per second, the double search space takes about $10^{30.2}$ years and the single permutation about
$10^{14.0}$ years. All of these are far beyond exhaustive search, so the reduction of the double scheme is not
operationally decisive. It does show that the chirp rate and the permutation domain are coupled design choices,
since $c_1$ is fixed by the full-diversity condition of \gls{AFDM}~\cite{bemani2023affine}, which makes it rational.

\subsection{PAPR and Out-of-Band Emission}
\label{sec:res_papr}

\begin{figure}[!tb]
    \centering
    \setlength{\figurewidth}{0.84\columnwidth}%
    \setlength{\figureheight}{3.3cm}%
    \input{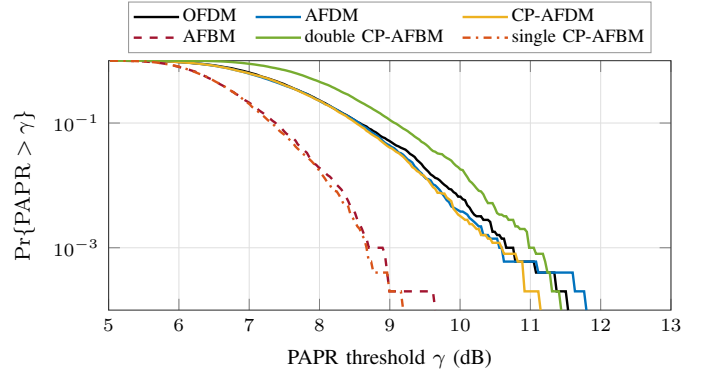}
    \caption{PAPR CCDF, $N=64$, 4-QAM, common $\times4$ oversampling, $5{,}000$ trials. The single CP-AFBM permutes
    only the $c_2$ chirp.}
    \label{fig:papr_performance}
\end{figure}

Fig.~\ref{fig:papr_performance} shows the \gls{PAPR} \gls{CCDF}, with every waveform measured on a common $\times4$
oversampling grid. At a \gls{CCDF} of $10^{-2}$, unpermuted \gls{AFBM} reaches $8.28$~dB against $9.81$~dB for
\gls{OFDM} and $9.65$~dB for \gls{AFDM}. Permuting $\bm{\lambda}_{c_1,L}$ costs that gain back: the double
CP-\gls{AFBM} reaches $10.20$~dB, $1.9$~dB above unpermuted \gls{AFBM} and at the level of \gls{OFDM}. The sequence
$\bm{\lambda}_{c_1,L}$ is the chirp whose quadratic phase progression flattens the envelope in \gls{DAFT}
spreading~\cite{tao2025affine}, and permuting it destroys that progression. The single permutation does not touch
it and reaches $8.18$~dB, which is not higher than unpermuted \gls{AFBM}. The comparison with \gls{OFDM} and \gls{AFDM},
whose blocks span $N$ samples against the $ON+\tfrac{N}{2}(K-1)$ of an \gls{AFBM} frame, is conservative for
\gls{AFBM}, since the maximum of a longer record is stochastically larger. The $1.9$~dB figure compares records of equal length.

\begin{figure}[!t]
    \centering
    \setlength{\figurewidth}{0.84\columnwidth}%
    \setlength{\figureheight}{3.3cm}%
    \input{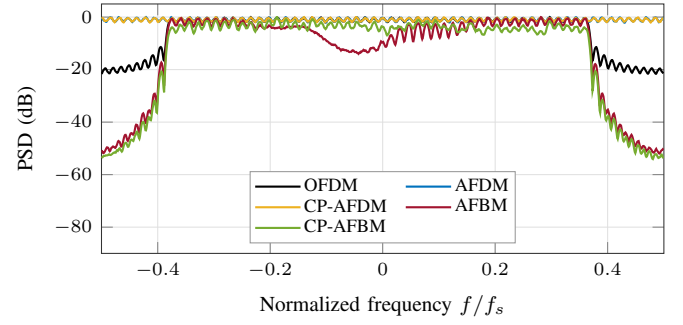}
    \caption{Out-of-band emission. OFDM is band-limited to the AFBM occupied bandwidth.}
    \label{fig:psd_performance}
\end{figure}

Fig.~\ref{fig:psd_performance} shows the power spectral density. \gls{OFDM} is loaded on only $P$ of its $N$
subcarriers so that it occupies the same band as \gls{AFBM}, whereas \gls{AFDM} and CP-\gls{AFDM} have no
frequency-containment stage and fill the band. \gls{AFBM} attains $-49.5$~dB against $-20.3$~dB for band-limited
\gls{OFDM}, and CP-\gls{AFBM} attains $-51.5$~dB. The permutation leaves the spectral containment intact because
the occupied band is fixed by the zero-padding of \eqref{eq:Q_P} and by the prototype, which the keys do not touch.

\section{Conclusion}
\label{sec:conclusion}
We proposed a double chirp-permuted \gls{AFBM} scheme with a secret lexicographic permutation on each of the two
\gls{DAFT} stages, and a single-permutation variant with one key on the shared $c_2$ chirp. A wrong key enters as a
diagonal between two Fourier stages, which makes the eavesdropper's product dense, so the precoding compensation
fails and the filter-bank interference leaks into the detector. Over a doubly-dispersive channel, an eavesdropper
with any wrong key of the double scheme has a \gls{BER} between $0.47$ and $0.51$ and a rate below $0.09$~bit/symbol,
even with perfect \gls{CSI} and co-location with the legitimate receiver, and blind carrier-phase recovery does not
improve this. The search space is $10^{93.3}$ against a key space of $10^{96.5}$, because the full-diversity condition
makes $c_1$ rational. The cost is a $1.9$~dB increase in \gls{PAPR}, and the spectral containment is unchanged.
The single permutation needs one key and adds no \gls{PAPR}, but a blind phase attack lowers the eavesdropper's \gls{BER}
to between $0.27$ and $0.33$. Future work includes selecting the chirp rate and the permutation domain jointly to recover the full
$L!\times P!$ key space, and making the prototype filter a second layer of secret parameters.

\balance

\selectlanguage{english}
\bibliographystyle{IEEEtran}
\bibliography{biblio.bib, references.bib}

\end{document}